\documentclass[superscriptaddress,aps,journal=prx,preprint]{revtex4-2}
\usepackage{graphicx}
\usepackage{epstopdf}
\usepackage{bm,amsmath,amssymb} 
\usepackage{wasysym}
\usepackage{physics}
\usepackage{gensymb}
\usepackage{color, soul} 
\usepackage{dcolumn}   
\usepackage{multirow}

\usepackage{lineno}
\begin{document}
\title{Skyrmion nucleus resolves the Landauer paradox}
\author{Jiyuan Yang}
\thanks{These authors contributed equally}
\affiliation{Department of Physics, School of Science, Westlake University, Hangzhou 310030, China}
\author{Denan Li}
\thanks{These authors contributed equally}
\affiliation{Department of Physics, School of Science, Westlake University, Hangzhou 310030, China}
\author{Shi Liu}
\email{liushi@westlake.edu.cn}
\affiliation{Department of Physics, School of Science, Westlake University, Hangzhou 310030, China}
\affiliation{Institute of Natural Sciences, Westlake Institute for Advanced Study, Hangzhou 310024, China}

\begin{abstract}
Ferroelectric switching is among the most fundamental and widely studied symmetry-breaking processes in physics. However, 
the Landauer paradox asserts that an ideal single-domain ferroelectric should be kinetically unswitchable, because a three-dimensional reversed nucleus would carry a prohibitive depolarization penalty, leading to unphysically large coercive field (the electric field required to switch the polarization). Here we show that this long-standing paradox is resolved by topology. In defect-free PbTiO\(_3\), large-scale molecular dynamics simulations reveal that the intrinsic critical nucleus is a three-dimensional polar skyrmion. Its continuous polarization rotation and N\'eel-type boundary walls self-compensate bound charge, suppressing the depolarization energy by orders of magnitude and bringing the predicted coercive field into agreement with experiment. 
This topological nucleus further overturns the 60-year-old Janovec--Kay--Dunn law. We derive an analytical thickness-dependent coercive-field law governed by field-induced softening of the parent ferroelectric state and controlled by a single material descriptor. This framework captures diverse experimental trends across ferroelectric families and recasts the Kay--Dunn exponent of \(-2/3\) as an effective finite-window behavior. These results establish polar topology as an organizing principle for intrinsic switching and suggest that hidden topological transition states may broadly shape nonequilibrium phase transformations traditionally understood as Landau-type symmetry breaking.
\end{abstract}

\maketitle
\newpage
\date{\today}

Ferroelectrics are materials with a switchable spontaneous electric polarization, a property that underpins a wide range of applications including nonvolatile memories and logic devices~\cite{Scott07p954, Scott89p1400}. Given its fundamental importance and technological relevance, ferroelectric switching might be expected to be a well-understood process. Yet a central question in polarization reversal remains unresolved, as captured by the famous Landauer paradox~\cite{Landauer57p227}: how can a uniformly polarized single-domain ferroelectric switch under an external electric field when classical nucleation theory predicts a prohibitively large energy cost for forming a three-dimensional (3D) reversed nucleus? 
At typical experimental coercive fields (the electric-field strength required to reverse the polarization), the estimated barrier exceeds \(10^8 k_{\rm B} T\), where \(k_{\rm B}\) is the Boltzmann constant and \(T\) is the temperature. Such an enormous barrier would make an ideal single-domain ferroelectric kinetically unswitchable, in direct contradiction with experiments.
This situation is analogous to the absence of raindrop formation in a perfectly dust-free atmosphere, where water vapor can remain supersaturated because nucleation is kinetically inhibited~\cite{Wyslouzil16p211702}. Over the years, this contradiction has been rationalized mainly through extrinsic mechanisms, including defect-assisted heterogeneous nucleation~\cite{Jesse08p209}, nucleation at surfaces or electrodes~\cite{Gerra05p107602}, ferroelastic or ferroelectric domain-wall intersections~\cite{Gao13p2791}, and charge injection or screening~\cite{Jiang09p024119}. This view leads to the unsettling implication that an ideal ferroelectric is not intrinsically switchable.
Atomistic and phase-field studies have largely focused on two-dimensional nucleation at pre-existing interfaces, especially domain walls, where switching is inherently easier~\cite{Choudhury08p162901,Shin07p881,Liu16p360}. How a 3D nucleus forms in a defect-free single domain, the problem closest to Landauer’s original formulation, has remained unresolved for almost 70 years.

Meanwhile, experiments show that the coercive field \(\mathcal{E}_c\) follows one of the most celebrated scaling laws in ferroelectrics, the Janovec--Kay--Dunn (JKD) law~\cite{Janovec58p3,Kay62p2027}. This law has been observed in a wide range of material systems, including perovskites~\cite{xu18p4736,Jiang22p779}, fluorites~\cite{Lyu19p220,Schroeder22p653}, and polymer ferroelectrics~\cite{dawber03pL393,Ducharme00p175}, for film thicknesses ranging from \(5~\mathrm{nm}\) to \(100~\mu\mathrm{m}\)~\cite{Dawber05p1083,CHANDRA04p7}. It states that \(\mathcal{E}_c \propto d^{-2/3}\), where \(d\) is the film thickness. However, the semiempirical JKD law was originally derived within Landauer's framework. This creates a fundamental inconsistency: a theory that appears to predict that ferroelectrics are not switchable is also used to explain how the coercive field varies with film thickness. Resolving the Landauer paradox while explaining the observed scaling behavior is central to a fundamental understanding of ferroelectric switching. 

In this work, using ferroelectric PbTiO$_3$ as a prototypical example, we perform large-scale molecular dynamics (MD) simulations to uncover the microscopic pathway of 3D nucleation during polarization reversal. By employing a deep-neural-network force field trained on first-principles data  and an enhanced sampling technique~\cite{Sun18p085703,Yang25p021042}, we determine the critical nucleus directly from MD simulations without prescribing its final shape. In contrast to the uniformly reversed nucleus assumed in the Landauer model, the critical nucleus is a polar skyrmion with a swirling polarization texture. This nontrivial topology produces a nearly self-compensated bound-charge distribution and enhanced dielectric screening, which together strongly suppress the depolarization field and remove the dominant energetic penalty responsible for the Landauer paradox. For a \(20~\mathrm{nm}\) thin film of PbTiO$_3$ at room temperature, the predicted coercive field is about \(0.3~\mathrm{MV/cm}\), in good agreement with experimental observations. This result highlights that a standard Landau-type symmetry-breaking process can proceed kinetically through a topological transition state.

We further construct an analytical skyrmion-nucleus model, which quantitatively predicts the thickness-dependent coercive field. Challenging the JKD law, the model shows that the \(\mathcal{E}_c\)--\(d\)
relation is a crossover, evolving from a spinodal-instability regime with an
effective exponent of about \(-0.55\) to a thick-film limit with an exponent of
\(-1\). The celebrated \(-2/3\) exponent appears only as an effective fit over a finite thickness window. Remarkably, the shape of the scaling relation is controlled by a single material descriptor, specifically \(P_s/K_{\mathrm{loc}}\), where \(P_s\) is the spontaneous polarization and \(K_{\mathrm{loc}}\) measures the depth of the ferroelectric double-well potential. The same framework extends to representative ferroelectric families, including Pb(Zr,Ti)O$_3$, BaTiO$_3$, KNbO$_3$, and polyvinylidene fluoride, providing a unified microscopic description of material-dependent coercive-field scaling.

\section*{Revisit the Landauer model}

We begin by revisiting the central assumptions of the Landauer model~\cite{Landauer57p227}. Consider a single-domain ferroelectric with remanent polarization \(P_s\) along the \(z\)-axis. The model assumes that a switched nucleus with polarization \(-P_s\) is separated from the parent domain with \(+P_s\) by a sharp interface (see Fig.~S3). The nucleation energy includes three main contributions: the volume term \(\Delta U_{\rm V}\) from the polarization–electric-field coupling, the interfacial energy \(\Delta U_{\rm I}\) due to the formation of the nucleus interface, and the depolarization energy \(\Delta U_{\rm D}\) arising from bound charges associated with the polarization discontinuity at the interface. Following the derivation of Landauer~\cite{Landauer57p227}, for a nucleus with lateral size \(l_x\) and vertical size \(l_z\), the energy can be expressed schematically as
$\Delta U_{\rm nuc} = \Delta U_{\rm V} + \Delta U_{\rm I} + \Delta U_{\rm D} =  -g_{\rm v} P_s \mathcal{E} l_x^2 l_z + g_{\rm i}\sigma l_x l_z + g_{\rm d}{l_x^4}/{l_z},
$
where \(g_{\rm v}\), \(g_{\rm i}\), and \(g_{\rm d}\) are geometry-dependent coefficients and \(\sigma\) is the domain-wall energy. Minimizing \(\Delta U_{\rm nuc}\) with respect to \(l_x\) and \(l_z\) yields the critical dimensions \(l_x^*\) and \(l_z^*\) (see Supplemental Sect.~III for details).

A key feature of the Landauer model is the electrostatic depolarization cost \(\Delta U_{\rm D}\). Across the top and bottom boundaries of the nucleus, the polarization is assumed to change abruptly from \(+P_s\) in the parent domain to \(-P_s\) in the switched region. By Gauss's law, this discontinuity produces a large bound-charge density, generating a strong depolarization field around the nucleus and consequently a substantial electrostatic energy penalty. To reduce this cost, the nucleus becomes highly elongated, with \(l_z^* \gg l_x^*\), to reduce the relative area of the charged interfaces. In this limit, the critical nucleus height scales as \(l_z^* \propto \sigma P_s^{-3/2}\mathcal{E}^{-3/2}\). If the coercive field \(\mathcal{E}_c\) is estimated by the condition \(l_z^* \approx d\), where \(d\) is the film thickness, inverting this relation gives the JKD law, \(\mathcal{E}_c \propto d^{-2/3}\) (see Supplemental Sect.~IV).
However, the expression for \(\Delta U_{\rm D}\) follows directly from the \textit{ad hoc} assumption of an abrupt polarization profile at the nucleus boundary. Resolving the paradox requires a first-principles-based determination of the polarization profile of a 3D critical nucleus.

\section*{A skyrmion nucleus for polarization switching}

Determining the critical nucleus is challenging because it is a short-lived saddle-point configuration. We address this challenge using large-scale MD simulations of defect-free PbTiO\(_3\) with a first-principles-trained machine-learning force field. Critical nuclei are identified over a wide range of electric fields using a revised persistent-embryo method (\(r\)PEM).
In \(r\)PEM, a small switched domain embedded in the parent domain is introduced as an embryo. A biasing potential stabilizes this embryo at the early stage and prevents its premature collapse. The bias is then gradually removed, allowing the embryo to evolve toward the saddle point. The bias vanishes completely while the nucleus remains smaller than the critical size, so configurations near the saddle-point region are sampled without bias. Details of the machine-learning model and \(r\)PEM are provided in the Methods.

\begin{figure}
	\begin{center}
		\includegraphics[width=1.0\textwidth]{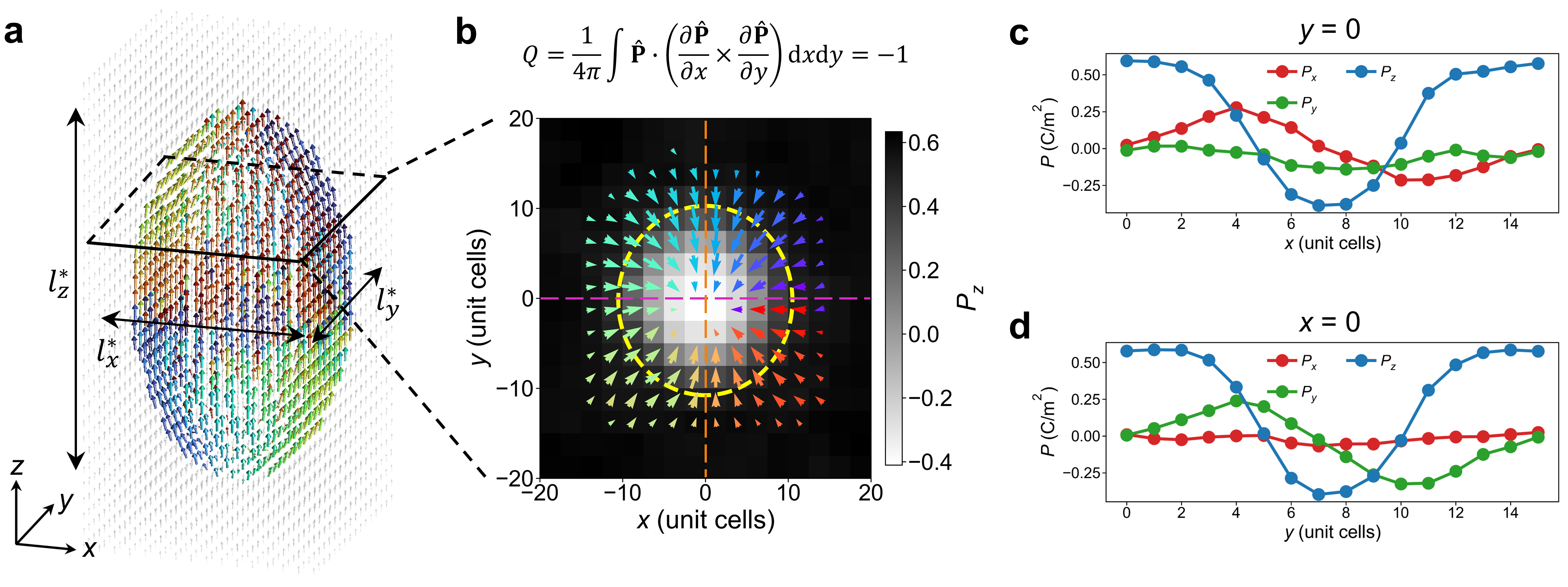}
	\end{center}
\caption{\textbf{Skyrmion-like critical nucleus during ferroelectric switching.}
\textbf{a}, Three-dimensional polarization texture of the critical nucleus obtained using MD-based \(r\)PEM at 300~K under a downward electric field of \(\mathcal{E}=0.55\)~MV/cm in single-domain PbTiO\(_3\) with polarization initially pointing along \(+z\). The lateral and vertical dimensions of the nucleus are denoted by \(l_x^*\), \(l_y^*\), and \(l_z^*\), respectively.
\textbf{b}, Polarization texture in the \(xy\) cross-section indicated by the boxed region in \textbf{a}. Arrows represent the in-plane polarization components, and the background color denotes the out-of-plane component \(P_z\). The continuous polarization rotation gives a topological charge of \(Q=-1\), identifying the nucleus as a polar skyrmion. The in-plane polarization points radially toward the nucleus center, demonstrating the convergent N\'eel-type boundary of the skyrmion nucleus.
\textbf{c}, Polarization components along an \(x\)-line profile in the \(y=0\) plane. The reversal of \(P_z\) is accompanied by a finite \(P_x\) and a nearly vanishing \(P_y\) at the wall.
\textbf{d}, Polarization components along a \(y\)-line profile in the \(x=0\) plane. The reversal of \(P_z\) is accompanied by a finite \(P_y\) and a nearly vanishing \(P_x\).}
 \label{Nucleus}
\end{figure}

At 300~K and under an applied electric field \(\mathcal{E}=0.55\)~MV/cm along the \(-z\) axis, the MD-based \(r\)PEM approach reveals that  the 3D critical nucleus is a polar skyrmion (Fig.~\ref{Nucleus}a). The occurrence of nucleation in a single-domain crystal at this relatively low field already shows that the Landauer model misses essential physics. The nucleus has lateral dimensions \(l_x^*=l_y^*=3.2\)~nm and a vertical dimension \(l_z^*=8.0\)~nm. This corresponds to an anisotropy ratio \(l_z^*/l_x^* \approx 2.5\), much smaller than the slender-nucleus limit \(l_z^* \gg l_x^*\) predicted by the Landauer model. 

Most importantly, the critical nucleus is characterized by smoothly rotating local polarizations across its boundary, as shown in the top view of the polarization profile for a layer cutting through the nucleus (Fig.~\ref{Nucleus}b).  The in-plane polarization component is normal to the wall and points radially toward the nucleus center, which is the defining feature of a N\'eel-type wall~\cite{Nataf20p634}. This N\'eel-type character is examined using line profiles across the nucleus boundary (Figs.~\ref{Nucleus}c and \ref{Nucleus}d). Along the line \(y=0\), the out-of-plane polarization \(P_z\) changes sign smoothly across the wall, while the in-plane component \(P_x\) becomes finite inside the wall and \(P_y\) remains nearly zero. Along the line \(x=0\), \(P_z\) again reverses continuously, \(P_y\) develops a finite value at the wall center, and \(P_x\) remains nearly zero. 

We further quantify the topology of this polarization texture by calculating the topological charge in the layer shown in Fig.~\ref{Nucleus}b:~$
Q=\frac{1}{4\pi}\int \hat{\mathbf P}\cdot
\left(
\frac{\partial \hat{\mathbf P}}{\partial x}
\times
\frac{\partial \hat{\mathbf P}}{\partial y}
\right)\dd x \dd y,
$
where \(\hat{\mathbf P}\) is the normalized local polarization. This calculation gives \(Q=-1\), confirming that the critical nucleus is a polar skyrmion. 
The emergence of this topological nucleus reveals a fundamental and surprising distinction between the thermodynamic and kinetic descriptions of ferroelectric switching. 
Thermodynamically, ferroelectric switching is usually viewed within Landau-type theory as polarization reversal between two symmetry-related polar states through a high-symmetry paraelectric configuration~\cite{Landau36p840, Lines77}. The kinetic pathway identified here is qualitatively different: switching proceeds through the nucleation of a topologically nontrivial polarization texture. Thus, a phase transformation traditionally understood as Landau-type symmetry breaking can be controlled kinetically by a hidden topological transition state.

\section*{Topological suppression of depolarization energy}

We next use the Landauer expression for a slender nucleus as a reference estimate to explain how the skyrmion polarization texture suppresses the depolarization energy \(\Delta U_\mathrm{D}\). This energy can be written as~\cite{Landauer57p227}
\begin{equation}
\Delta U_\mathrm{D} \approx 
\frac{\pi \varrho_{\mathrm{b}}^2 \delta^2 l_x^4}
{12\epsilon_0  \epsilon_a l_z}
\left[
\ln \left( \frac{2l_z}{l_x} \sqrt{\frac{\epsilon_a}{\epsilon_c}} \right) - 1
\right],
\label{Ud}
\end{equation}
where \(\delta\) is the domain-wall width, \(\varrho_{\mathrm{b}}\) is the effective bound-charge density at the nucleus boundary, and \(\epsilon_a\) and \(\epsilon_c\) are the in-plane and out-of-plane dielectric constants, respectively. 
Because the Landauer model assumes a sharp interface with a domain-wall width of one unit cell,
\(\delta \approx a = 3.92~\text{\AA}\), the bound-charge density is $\varrho_{\mathrm{b}} = {2P_s}/{a}
= 3.6 \times 10^9~\text{C m}^{-3}$, using $P_s=0.7$~C/m$^2$ at room temperature.
The dielectric response of the nucleus is also assumed to be bulk-like, with
\(\epsilon_a \approx \epsilon_c \approx 130\) for PbTiO\(_3\)~\cite{Kakuta95p5341}. 
For an external field of \(0.55~\mathrm{MV/cm}\), representative of typical experimental coercive fields in thin films at \(T=300~\mathrm{K}\), and using \(\sigma=0.469~\mathrm{J/m^2}\) for head-to-head/tail-to-tail charged domain walls, the Landauer model predicts a slender critical nucleus with \(l_x^*=24~\mathrm{nm}\) and \(l_z^*=256~\mathrm{nm}\) (an order of magnitude larger than the film thickness). The corresponding depolarization energy is \(\Delta U_{\rm D}^* \approx 2.9 \times 10^5\,k_{\rm B}T\), and the nucleation barrier is \(\Delta U_{\rm nuc}^* = 5.7\times10^5\,k_{\rm B}T\), implying a vanishingly small probability of nucleation (see Supplemental Sect.~III for additional details).

\begin{figure}
	\begin{center}
		\includegraphics[width=1.0\textwidth]{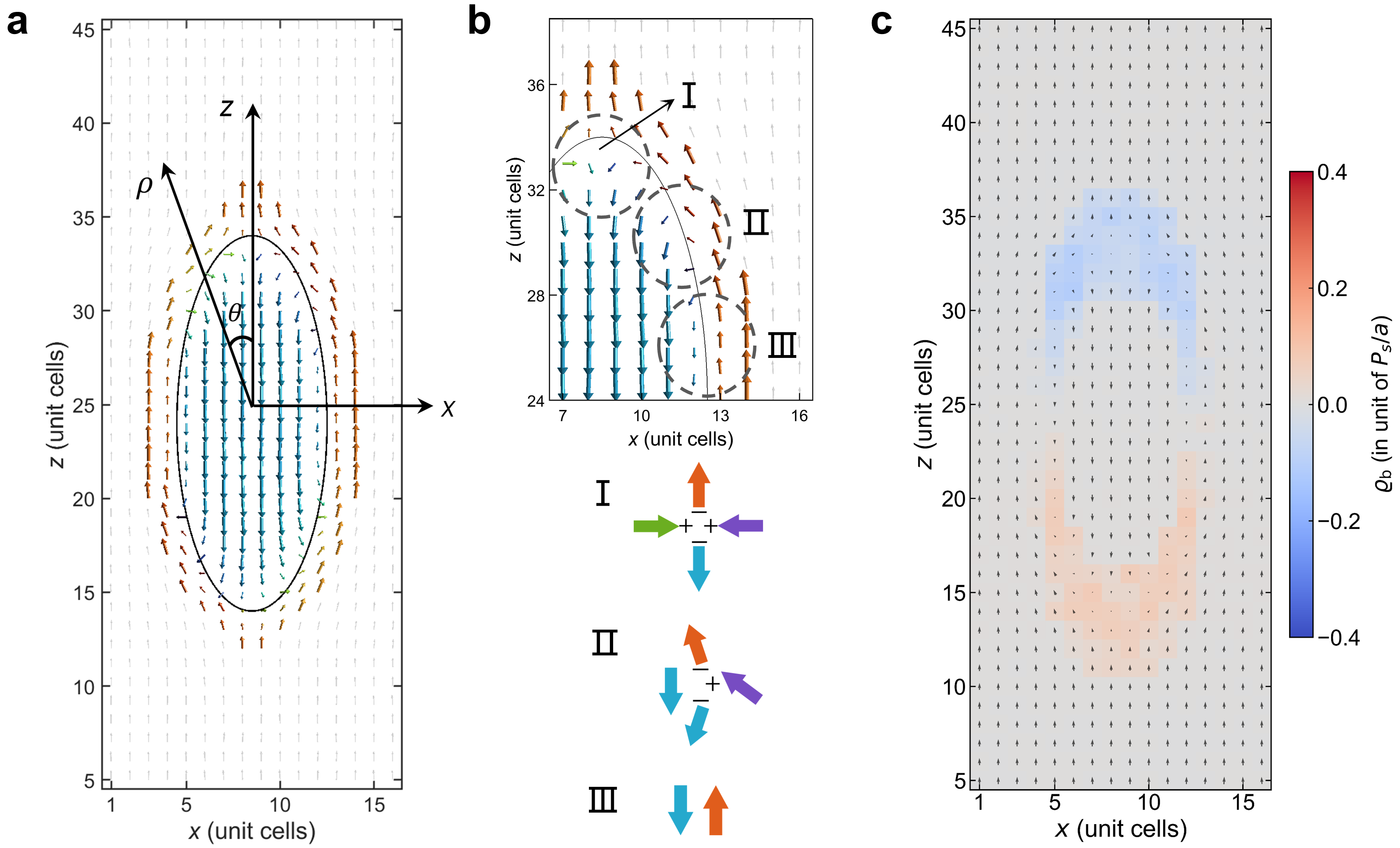}
	\end{center}
    \caption{\textbf{Suppression of bound-charge density by the continuously rotating polarization texture.}
\textbf{a}, Polarization texture in the central $xz$ cross-section of the critical skyrmion nucleus. The ellipsoidal boundary is parameterized by the radial coordinate $\rho$ and polar angle $\theta$ to evaluate the anisotropic interfacial energy.
\textbf{b}, Enlarged view of the upper boundary of the nucleus. Three representative regions are distinguished according to the configuration of the out-of-plane polarization: region I corresponds to a nominally strongly charged tail-to-tail configuration, region II to a partially compensated configuration, and region III to a charge-neutral configuration. The schematics below illustrate the local polarization arrangements in these regions. In regions I and II, the bound charge expected from the out-of-plane polarization gradient is largely compensated by the convergent in-plane polarization components.
\textbf{c}, Spatial distribution of the net bound-charge density,
$\varrho_\mathrm{b}=-\boldsymbol{\nabla}\cdot\mathbf{P}$, calculated from the full 3D polarization field. The color scale is given in units of $P_s/a$, where $P_s$ is the spontaneous polarization and $a$ is the lattice constant. The weak residual charge at the nucleus boundary demonstrates the near-self-compensation produced by the skyrmion polarization texture.}
 \label{UD_and_Ec}
\end{figure}

For the critical skyrmion nucleus, both assumptions in the Landauer model are fundamentally altered. First, continuous polarization rotation strongly reduces the effective bound-charge density. As shown in Fig.~\ref{UD_and_Ec}b, if only the out-of-plane polarization gradient \(\partial P_z/\partial z\) is considered, the nucleus boundary can be divided into three representative regions: region~\(\mathrm{I}\), a nominally strongly charged tail-to-tail configuration; region~\(\mathrm{II}\), a partially charged tail-to-tail configuration; and region~\(\mathrm{III}\), a nearly charge-free region..
The actual local bound-charge density, however, is determined by the full 3D divergence,
$
\varrho_{\mathrm{b}}
= -\nabla \cdot \mathbf{P} = 
-\left(
\frac{\partial P_x}{\partial x}
+
\frac{\partial P_y}{\partial y}
+
\frac{\partial P_z}{\partial z}
\right).
$
In region~\(\mathrm{I}\), the substantial charge expected from \(\partial P_z/\partial z\) is largely canceled by the in-plane divergence, \(\partial P_x/\partial x+\partial P_y/\partial y\). This cancellation is intrinsic to the skyrmion geometry: the out-of-plane tail-to-tail configuration is accompanied by an in-plane head-to-head arrangement, as schematically shown in the bottom panel of Fig.~\ref{UD_and_Ec}b.
As a result, boundary regions that would be highly charged in the Landauer picture carry only a small bound charge. A similar, although weaker, compensation occurs in region~\(\mathrm{II}\). Quantitatively, our numerical calculations show that the average bound-charge density is reduced to \(\varrho_{\mathrm{b}} \approx 0.047P_s/a =  8.4 \times 10^{7}\,\mathrm{C\,m^{-3}}\) (see Fig.~\ref{UD_and_Ec}c), nearly two orders of magnitude lower than the sharp-boundary value of \(3.6 \times 10^{9}\,\mathrm{C\,m^{-3}}\).

Second, the skyrmion texture strongly enhances dielectric screening because dipoles near the nucleus boundary are highly susceptible to rotation, leading to effective dielectric constants that exceed the bulk values~\cite{Wojdel14p247603, Liu17p094102}. Since \(\epsilon_a\) appears in the denominator of Eq.~(\ref{Ud}), this enhanced screening further suppresses \(\Delta U_{\mathrm{D}}\). We therefore use \(\epsilon_a \approx \epsilon_c \approx 800\), consistent with previous experimental observations in superlattices supporting polar skyrmions in PbTiO\(_3\)~\cite{Das21p194}. Together, the reduced bound-charge density and enhanced dielectric screening lower $\Delta U_{\rm D}$ defined in Eq.~(\ref{Ud}) to \(0.42\,k_{\mathrm{B}}T\) for the critical skyrmion nucleus (\(l_x^*=3.2~\mathrm{nm}\), \(l_z^*=8.0~\mathrm{nm}\)) shown in Fig.~\ref{Nucleus}a, about six orders of magnitude below the Landauer prediction.

\section*{Skyrmion nucleus model}

Because the skyrmion nucleus has a nearly vanishing depolarization contribution, its energy relative to the single-domain state can be written as
\begin{equation}
\Delta U_{\mathrm{nuc}}
=
\Delta U_{\mathrm{V}}
+
\Delta U_{\mathrm{I}}
=
-\frac{\pi}{3}l_x^2 l_z P_s\mathcal{E}
+
2\pi\int_0^\pi
\sqrt{
(\sigma_x^{\text{N\'{e}el}})^2 \sin^2\theta
+
(\sigma_z^{\text{N\'{e}el}})^2 \cos^2\theta
}
\left(\frac{l_\theta}{2}\right)^2
\sin\theta\,\mathrm{d}\theta .
\label{SkrNuc}
\end{equation}
Detailed derivations are given in  Supplemental Sect.~VII.
The volume term \(\Delta U_{\mathrm{V}}\) is the same as that for an ellipsoidal nucleus of the same dimensions with uniformly reversed polarization, as the rotational polarization components cancel in the volume integral by symmetry.
The dominant energy cost is the interfacial energy, \(\Delta U_{\mathrm{I}}\), associated with the formation of N\'eel-type walls at the nucleus boundary. As shown in Fig.~\ref{SkrNuc}a, we use spherical coordinates \((\rho,\theta)\) to parameterize the ellipsoidal surface, with \(l_\theta=l_xl_z/\sqrt{(l_x\cos\theta)^2+(l_z\sin\theta)^2}\). The interfacial contribution is then \(\Delta U_{\mathrm{I}}=\iint \sigma_\rho^\text{N\'{e}el}\,\mathrm{d}S\), where \(\sigma_\rho^{\text{N\'{e}el}}\) is the energy of a N\'eel-type wall along the radial direction specified by \(\rho\) and \(\theta\). We approximate the anisotropic \(\sigma_\rho^{\text{N\'{e}el}}\) using two representative values, \(\sigma_x^{\text{N\'{e}el}}\) and \(\sigma_z^{\text{N\'{e}el}}\), corresponding to \(180^\circ\) N\'eel-type walls perpendicular to the \(x\) and \(z\) axes, respectively; the resulting expression using \(\sigma_x^{\text{N\'{e}el}}\) and \(\sigma_z^{\text{N\'{e}el}}\) is given in Eq.~(\ref{SkrNuc}). A similar expression has been rigorously derived for Ising-type walls in ferroelectric hafnia~\cite{Yang25p021042}.

We emphasize that the analytical skyrmion-nucleus model defined in Eq.~(\ref{SkrNuc}) depends only on three intrinsic material parameters: \(\sigma_x^{\text{N\'eel}}\), \(\sigma_z^{\text{N\'eel}}\), and \(P\). A subtle but important point is that these parameters must be evaluated at the relevant finite field and finite temperature, \(\mathcal{E}\) and \(T\). This can be done straightforwardly by scaling their zero-Kelvin, zero-field values, which can be readily computed once the Curie temperature is known (see Methods). The critical size \((l_x^*, l_z^*)\) is then obtained from the stationary condition of \(\Delta U_{\rm nuc}\), and the corresponding energy defines the nucleation barrier \(\Delta U_{\rm nuc}^*\). As shown in Fig.~\ref{Unuc_and_Ec}a, the critical dimensions \(l_x^*\) and \(l_z^*\) predicted by the skyrmion-nucleus model over a range of field strengths are in excellent agreement with those obtained directly from MD-based \(r\)PEM simulations, validating the accuracy and robustness of the nucleation model. By contrast, the Landauer model, as expected, yields an unphysically large and slender nucleus with \(l_z^* \gg l_x^*\). Figure~\ref{Unuc_and_Ec}b shows the corresponding nucleation barriers, demonstrating that the skyrmion-nucleus model lowers the barrier by about three orders of magnitude compared with the Landauer prediction. 

We can further relate the field-dependent critical-nucleus size to the thickness-dependent coercive field of PbTiO\(_3\) thin films. Following the
assumption of Kay and Dunn, for a film of thickness \(d\), \(\mathcal{E}_c\) is the field at which \(l_z^* \approx d\). In this regime, ferroelectric switching is governed by nucleation-limited switching (NLS). The skyrmion-nucleus model predicts \(\mathcal{E}_c=0.3~\mathrm{MV/cm}\) for initiating polarization switching in 20-nm-thick PbTiO$_3$ thin films through 3D nucleation at 300~K,  comparable to experimental values of $\approx0.5$~MV/cm under the same conditions~\cite{Nishino20p10864}. This result resolves the Landauer paradox by revealing that a topologically nontrivial skyrmion nucleus provides a physically viable pathway for ferroelectric switching.

\section*{Origin of coercive-field scaling}

Resolving Landauer's paradox by showing that the depolarization energy is negligible seems to raise an apparent new puzzle. The classical JKD scaling, \(\mathcal{E}_c \propto d^{-2/3}\), which appears to hold across a wide range of ferroelectric materials, depends directly on the depolarization-energy term in the Landauer model. We now show that the skyrmion-nucleus model can nevertheless produce quantitatively similar scaling behavior, but with an important caveat: the apparent \(-2/3\) exponent is only an effective approximation.

Because the depolarization energy is negligible, the energy of a skyrmion nucleus can be expressed schematically as
$\Delta U_{\rm nuc} = -g_{\rm v} P\mathcal{E} l_z^3
+ g_{\rm i}\sigma l_z^2$.
The saddle-point condition gives
\begin{equation}
    l_z^* = \frac{2g_{\rm i}\sigma}{3g_{\rm v}\mathcal{E}P} .
    \label{lz}
\end{equation}
By setting $l_z^* \approx d$ at $\mathcal{E}=\mathcal{E}_c$, while treating polarization $P$ and domain-wall energy $\sigma$ as field-independent constants, Eq.~(\ref{lz}) would predict $\mathcal{E}_c \propto d^{-1}$ rather than the JKD scaling. The crucial point is that both \(P\) and \(\sigma\) depend on the applied field, written as \(P(\mathcal{E})\) and \(\sigma(\mathcal{E})\). We start with the zero-field spontaneous polarization \(P_s\) at finite temperature \(T\), so that thermal softening is already included (see Methods). Under a switching field opposite to \(P_s\), the polarization of the parent domain is softened to a field-dependent value \(P(\mathcal{E})\) before nucleation occurs, reflecting the intrinsic dielectric and piezoelectric response of the ferroelectric. The domain-wall energy also decreases as the parent domain softens.
From Landau--Ginzburg theory, it follows the characteristic scaling, $\sigma(P) = \sigma_{0} (P/P_s)^3$, where $\sigma_0=\sigma(P_s)$ is the zero-field domain-wall energy at $T$ (see Supplementary Sect.~VI for derivations). Using the softened domain-wall energy in Eq.~(\ref{lz}) gives
\begin{equation}
    l_z^* = \frac{2g_{\rm i}\sigma_0}{3g_{\rm v}\mathcal{E}P_s}
    \left(\frac{P}{P_s}\right)^2 .
    \label{sKrlz}
\end{equation}
where $P$ is the softened polarization value $P(\mathcal{E})$ at $\mathcal{E}$.
To determine \(P(\mathcal{E})\), we start from the electric enthalpy of a ferroelectric under an applied field and solve it perturbatively (see Methods). Substituting \(P(\mathcal{E})\) into Eq.~(\ref{sKrlz}) and imposing \(l_z^*=d\) at \(\mathcal{E}=\mathcal{E}_c\) yields
\begin{equation}
d = \frac{2g_\mathrm{i}\sigma_0}{3g_\mathrm{v}P_s\mathcal{E}_c}
\left[
1-\frac{1}{8}\frac{\mathcal{E}_c}{\mathcal{E}^*}
-\frac{3}{128}
\left(\frac{\mathcal{E}_c}{\mathcal{E}^*}\right)^2
\right]^2,
\label{d(E)}
\end{equation}
where \(\mathcal{E}^*=K_{\rm loc}/P_s\) is the characteristic softening field, and \(K_{\rm loc}\) is the depth of the ferroelectric double-well potential at $T$.

\begin{figure}
	\begin{center}
		\includegraphics[width=1.0\textwidth]{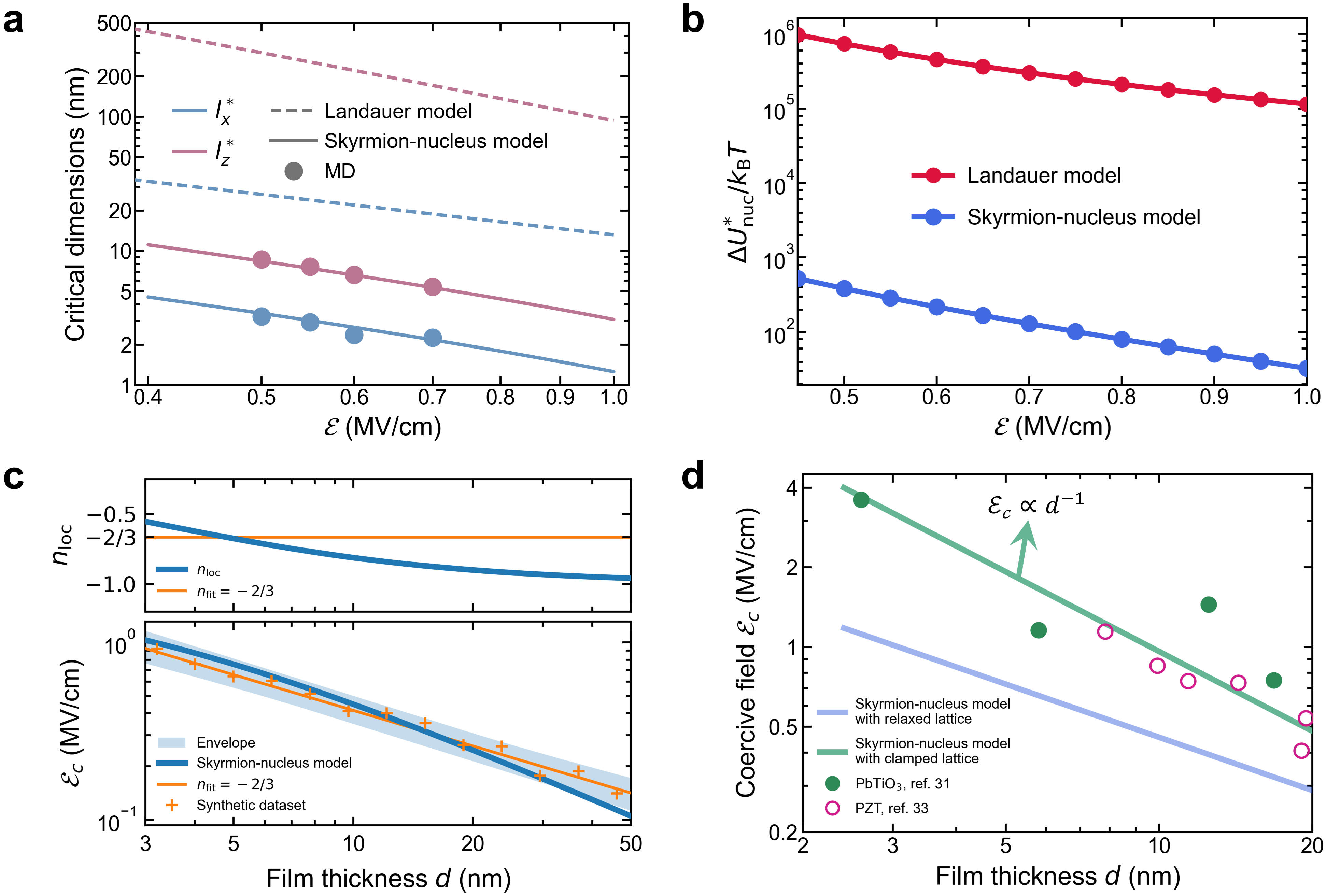}
	\end{center}
  \caption{\textbf{Landauer paradox resolved by skyrmion nucleation.}
\textbf{a}, Critical nucleus dimensions as functions of the applied field. Solid lines denote the skyrmion-nucleus model, circles denote MD-based \(r\)PEM results, and dashed lines denote the Landauer model for PbTiO\(_3\).
\textbf{b}, Nucleation barriers predicted by the skyrmion-nucleus and Landauer models.
\textbf{c}, Thickness-dependent coercive field, \(\mathcal{E}_c\), predicted by the skyrmion-nucleus model. The top panel shows a continuous crossover of the local exponent, $n_{\mathrm{loc}}=\mathrm{d}\ln\mathcal{E}_c/\mathrm{d}\ln d$, from \(-1\) in thick films to \(-0.55\) in ultrathin films. Datasets within the envelope around the exact skyrmion-nucleus-model curve can be fitted to a single power law with exponent \(-0.7<n_{\rm fit}<-0.6\) while maintaining \(R^2>0.9\).
\textbf{d}, Coercive-field scaling calculated with and without lattice clamping, compared with experimental data for PbTiO$_3$ and PZT thin films.}
 \label{Unuc_and_Ec}
\end{figure}

Equation~(\ref{d(E)}) is the central result for the coercive-field scaling. Figure~\ref{Unuc_and_Ec}c shows the \(\mathcal{E}_c\)--\(d\) relation for PbTiO$_3$ at 300~K, obtained by numerically inverting Eq.~\eqref{d(E)}. The resulting curve is not a simple universal power law, but two limiting behaviors can be derived  analytically. In the thick-film limit, the coercive field is small compared with the characteristic softening field, \(\mathcal{E}_c \ll \mathcal{E}^*\). The correction factor in the bracket of Eq.~(\ref{d(E)}) approaches unity, giving \(d \propto \mathcal{E}_c^{-1}\), or equivalently \(\mathcal{E}_c \propto d^{-1}\). 
Physically, this is the weak-field-softening regime: the applied field only weakly changes the parent-domain polarization, so \(P(\mathcal{E}_c)\approx P_s\) and both \(P\) and \(\sigma\) are nearly field independent, recovering the simple \(d^{-1}\) scaling discussed above.
In the opposite ultrathin-film limit, \(\mathcal{E}_c\)  approaches the maximum opposing field $\mathcal{E}^{\rm max}$ at which the polar state loses stability. This spinodal-like instability is determined by the unstable stationary point of the electric enthalpy, giving
\(\mathcal{E}^{\rm max}=\frac{8}{3\sqrt{3}}\,\mathcal{E}^*\).
Substituting this limiting field into Eq.~(\ref{d(E)}) gives a weaker local scaling, approximately \(\mathcal{E}_c \propto d^{-0.55}\) (see Supplementary Sect.~VIII).
Thus, as the film thickness decreases from the thick- to ultrathin-film regime, the local scaling exponent \(n_{\rm loc}=\mathrm{d}\ln \mathcal{E}_c/\mathrm{d}\ln d\) continuously evolves from \(-1\) toward \(-0.55\)., as shown in the top panel of Fig.~\ref{Unuc_and_Ec}c.

In practice, coercive-field measurements are performed over finite thickness windows and are inevitably influenced by extrinsic factors such as sample quality, defects, and electrode effects~\cite{liang26pe75841}. Consequently, experimental \(\mathcal{E}_c\)--\(d\) datasets often exhibit appreciable scatter~\cite{CHANDRA04p7}. Interestingly, such scatter is often invoked to regard experimentally measured exponents ranging from approximately \(-0.55\) to \(-1.0\) as broadly consistent with the JKD value of $-2/3$~\cite{xu18p4736,Nishino20p10864}.

We suggest that this same scatter can obscure the intrinsic curvature of the underlying
$\mathcal{E}_c$--$d$ relation and make it appear consistent with a
single power law. To illustrate this point, we consider data distributed
around the exact skyrmion-model curve and fit them to
$\mathcal{E}_c \propto d^{n_{\rm fit}}$, where $n_{\rm fit}$ is the
exponent extracted from a single-power-law fit. The coefficient of determination, \(R^2\), measures the fit quality, with \(R^2=1\) corresponding to a perfect power law. In Fig.~\ref{Unuc_and_Ec}c, we identify the region around the exact curve in which datasets can yield a JKD-like exponent, \(-0.7<n_{\rm fit}<-0.6\), while maintaining \(R^2>0.9\). This broad region shows that modest deviations from the intrinsic curve can mask its non-power-law character and produce apparent JKD scaling. As a representative example,
Fig.~\ref{Unuc_and_Ec}c shows a synthetic dataset with a mean absolute error of \(12\%\) relative to the theoretical prediction, comparable to typical experimental scatter, yet giving \(n_{\rm fit}=-0.67\). This suggests an apparent \(-2/3\) exponent can arise naturally from finite-window fitting in the presence of realistic experimental uncertainty.

The skyrmion-nucleus model also explains experimental coercive-field  data in PbTiO\(_3\) thin films that deviate strongly from JKD scaling. The model predicts \(\mathcal{E}_c\propto d^{-1}\) whenever field-induced polarization softening is weak. This occurs not only in the thick-film limit, where \(\mathcal{E}_c\) is small, but also in strongly lattice-clamped thin films, where the polarization remains close to \(P_s\) even under large applied fields.
As shown in Fig.~\ref{Unuc_and_Ec}d, we calculate the \(\mathcal{E}_c\)--\(d\) curve for PbTiO\(_3\) using parameters for a clamped lattice, with \(T_c=1300~\mathrm{K}\) computed from fixed-lattice MD simulations as the only modified parameter. The resulting curve lies above the relaxed-lattice prediction, exhibits a clear \(\mathcal{E}_c\propto d^{-1}\) scaling, and agrees well with two independent experimental datasets~\cite{Nishino20p10864,Pertsev03p3356}. The skyrmion-nucleus model thus resolves the Landauer coercive-field paradox and captures experimentally observed thickness scaling beyond JKD behavior.

\section*{Unified scaling across ferroelectric families}

Equation~(\ref{d(E)}) provides a fundamentally different interpretation of coercive-field scaling. The thickness dependence of \(\mathcal{E}_c\) is controlled by two fitting parameters: the inverse characteristic softening field, \(P_s/K_{\mathrm{loc}}=1/\mathcal{E}^*\), and the prefactor \(2g_{\mathrm{i}}\sigma_0/(3g_{\mathrm{v}}P_s)\). Thus, Eq.~(\ref{d(E)}) has the same number of fitting parameters as a conventional single-power-law fit, \(\mathcal{E}_c\propto d^n\), which contains a scaling exponent and an intercept. 

\begin{figure}
	\begin{center}
		\includegraphics[width=0.8\textwidth]{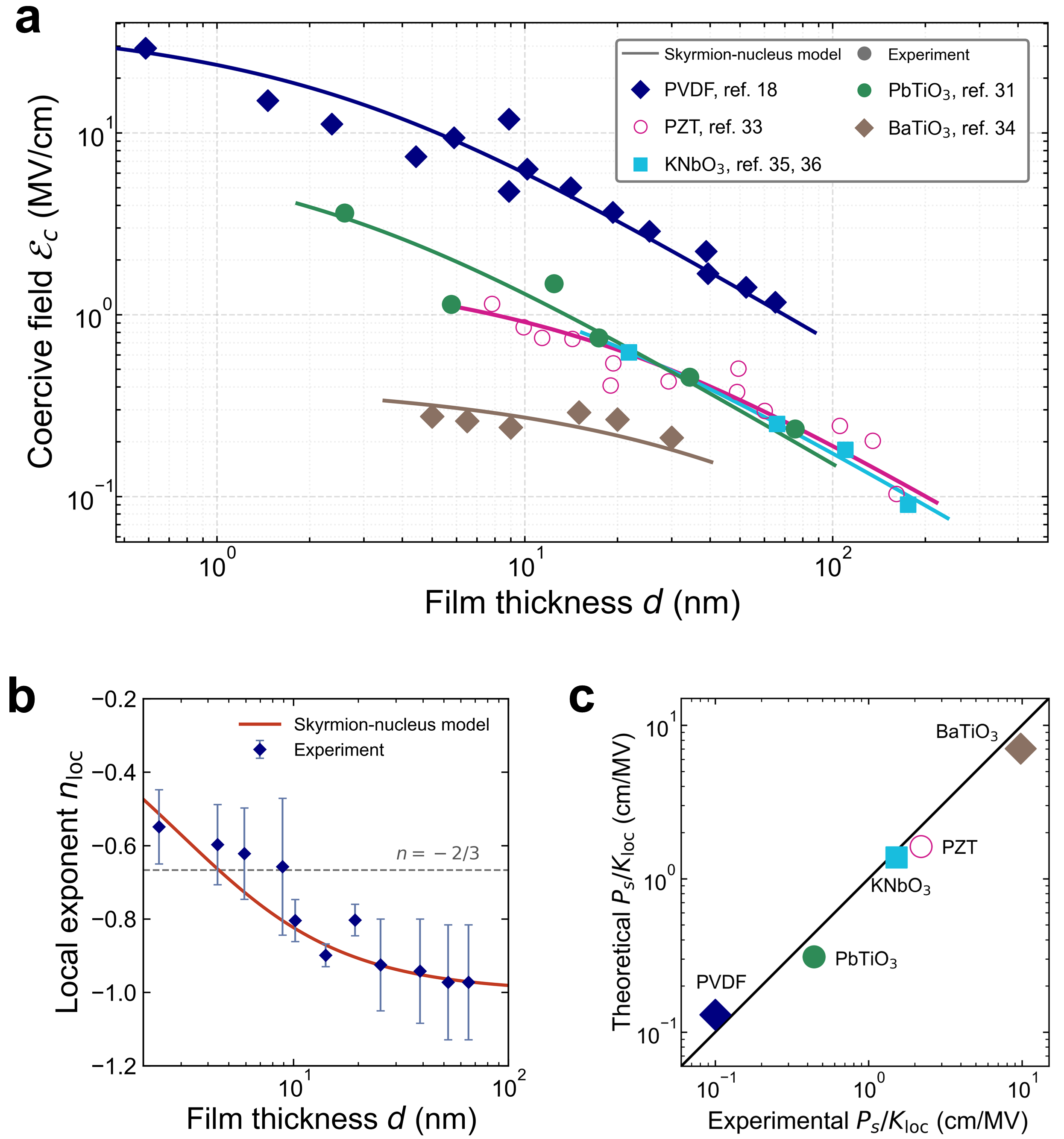}
	\end{center}
    \caption{\textbf{Unified description of coercive-field scaling across chemically distinct ferroelectrics.}
\textbf{a}, Experimental thickness-dependent coercive fields of PVDF, PbTiO\(_3\), PZT, BaTiO\(_3\), and KNbO\(_3\), compared with predictions of the skyrmion-nucleus model. Symbols denote experimental data, and solid curves denote fits using Eq.~(\ref{d(E)}).
\textbf{b}, Comparison between the local scaling exponent \(n_{\mathrm{loc}}\) predicted by the skyrmion-nucleus model and that extracted from PVDF experiments. At each experimental thickness, \(n_{\mathrm{loc}}\) is obtained by fitting the five neighboring data points, with error bars representing \(\pm 1\) standard error.
\textbf{c}, Comparison between independently estimated theoretical and experimental values of the intrinsic material descriptor \(P_s/K_{\mathrm{loc}}\). The diagonal line denotes exact agreement. The cross-material consistency supports \(P_s/K_{\mathrm{loc}}\) as a physically meaningful descriptor that controls coercive-field scaling.}
 \label{UniEc}
\end{figure}

We test Eq.~(\ref{d(E)}) using experimental coercive-field data for polyvinylidene fluoride (PVDF) and four representative perovskite ferroelectrics: PbTiO\(_3\), Pb(Zr,Ti)O\(_3\) (PZT), BaTiO\(_3\), and KNbO\(_3\)~\cite{dawber03pL393,Nishino20p10864,Pertsev03p3356,jo06p23,Luo23p16902,Hazra24p2408664} (Fig.~\ref{UniEc}a). These materials differ substantially in chemistry, spontaneous polarization, coercive-field magnitude, and accessible thickness range. A single-power-law analysis of these datasets gives widely varying exponents, as summarized in Table~S2, ranging from \(n=-0.55\) to \(n=-1.0\). The broad distribution already suggests that the thickness scaling is not governed by a universal power-law exponent. By contrast, Eq.~(\ref{d(E)}) captures both the magnitude and the curvature of the measured \(\mathcal{E}_c\)--\(d\) relations within the same skyrmion-nucleus framework, with fitting quality comparable to or better than that of the corresponding single-power-law fits. 

The advantage of Eq.~(\ref{d(E)}) becomes especially clear when examining the local scaling exponent,
\(n_{\mathrm{loc}}=\mathrm{d}\ln\mathcal{E}_c/\mathrm{d}\ln d\). As shown in Fig.~\ref{UniEc}b for PVDF, the experimentally extracted \(n_{\mathrm{loc}}\) is not constant. Instead, it varies appreciably with thickness and becomes more negative as the film thickness increases, decreasing from values close to \(-0.55\) in thinner films toward values approaching \(-1\) in thicker films. This trend is inconsistent with a fixed JKD exponent \(n=-2/3\), but is well captured by the prediction of Eq.~(\ref{d(E)}). 

Finally, a more stringent test is provided by the fitted values of \(P_s/K_{\mathrm{loc}}\). Unlike a phenomenological power-law exponent, \(P_s/K_{\mathrm{loc}}\) at finite temperature is a fundamental material descriptor which can be independently computed from first-principles calculations of the intrinsic ferroelectric energy landscape with proper thermal scaling (see Supplementary Sect.~IX). As shown in Fig.~\ref{UniEc}c, the values of \(P_s/K_{\mathrm{loc}}\) extracted from experimental coercive-field data agree well with our first-principles estimates across two orders of magnitude (see values in Table~S3). This agreement is nontrivial because the two values are obtained from entirely different sources: one from macroscopic switching measurements and the other from material energetics. It demonstrates that \(P_s/K_{\mathrm{loc}}\) surprisingly controls the \(\mathcal{E}_c\) scaling curve with film thickness.

\section*{Discussion}

The Landauer paradox has long exposed a fundamental inconsistency in our understanding of ferroelectric switching: classical nucleation theory appears to imply that an ideal single-domain ferroelectric should be kinetically unswitchable. Here we show that a ferroelectric possesses an intrinsic route around this bottleneck. The critical nucleus for polarization reversal takes the form of a three-dimensional polar skyrmion. Through continuous polarization rotation, the bound-charge density associated with reversal of the out-of-plane polarization is largely compensated by the divergence of the in-plane polarization. At the same time, the enhanced rotational susceptibility near the skyrmion boundary provides strong dielectric screening. As a result, the depolarization energy is suppressed by orders of magnitude and no longer dominates the nucleation barrier. The apparent impossibility of intrinsic switching is therefore resolved by topology.

This finding reveals a previously unrecognized role of ferroelectric topological structures. Topological polar textures have become a frontier in condensed matter physics, particularly following their stabilization in engineered oxide superlattices~\cite{Junquera23p025001,Das2019,Naumov04p737,pereira19peaau7023,gonccalves24p066802}. Our results show that their significance extends beyond equilibrium or metastable states. A polar skyrmion can also emerge transiently as the saddle-point configuration of a nonequilibrium transformation, controlling the kinetics of polarization reversal itself. In this sense, topology does more than enrich the landscape of possible ferroelectric states; it can also determine the kinetic pathway by which the system traverses that landscape.

The skyrmion-nucleus model also challenges the 60-year-old JKD law. The widely used scaling \(\mathcal{E}_c \propto d^{-2/3}\) has traditionally been linked to Landauer-type depolarization arguments. We find instead that the thickness dependence of the coercive field originates from field-induced softening of the parent ferroelectric state. The applied switching field reduces the polarization before nucleation, which modifies both the thermodynamic driving force and the domain-wall energy. Consequently, the resulting \(\mathcal{E}_c\)--\(d\) relation is a crossover whose local exponent evolves continuously with thickness. The JKD exponent emerges only as an effective finite-window approximation within this broader intrinsic scaling relation.

A central outcome of this work is that coercive-field scaling, despite being a complex macroscopic manifestation of switching kinetics, can be related quantitatively to a single intrinsic material descriptor. Specifically, the descriptor \(P_s/K_{\mathrm{loc}}\), which measures the susceptibility of the ferroelectric energy landscape to field-induced softening, controls the curvature of the thickness-scaling relation. Because this parameter can be obtained directly from first-principles calculations, our theory enables the prediction of thickness-dependent NLS-type coercive fields, a task that might otherwise appear beyond the scope of first-principles methods. 

In summary, our results establish topological nucleation as the microscopic mechanism that reconciles intrinsic ferroelectric switchability with classical nucleation theory and determines the scaling of dynamical properties such as the coercive field. Although ferroelectric switching has long been described as a Landau-type symmetry-breaking phenomenon, our findings reveal that the reversal proceeds through a topological transformation of the polarization texture. More broadly, this work points to hidden topological transition states as an organizing principle for symmetry-breaking phase transformations beyond ferroelectricity.

\section*{Methods}

\subsection*{Molecular Dynamics}

The molecular dynamics (MD) simulations were performed using a previously developed deep-potential (DP) model for Pb$_x$Sr$_{1-x}$TiO$_3$~\cite{Wu23p144102}. The DP model represents the total energy of a configuration as a sum of atomic energies predicted from local atomic environments using a deep neural network~\cite{Zhang18p143001}. The training database contains 19,119 Pb$_x$Sr$_{1-x}$TiO$_3$ configurations that sample different phases and their thermally distorted structures. The reference energies and atomic forces were obtained from density functional theory (DFT) calculations using the Vienna \textit{ab initio} Simulation Package (VASP) with the projector-augmented-wave method~\cite{Kresse96p11169,Kresse96p15} and the Perdew--Burke--Ernzerhof functional revised for solids (PBEsol)~\cite{Perdew08p136406}. A plane-wave energy cutoff of 800~eV and a \(k\)-point spacing of 0.3~\AA$^{-1}$ were used.  The DP model has a mean absolute error of 0.825~meV/atom for energies and 0.037~eV/\AA~for atomic forces. It has also been comprehensively validated against first-principles calculations and shown to reproduce key properties of PbTiO$_3$, including the phonon spectra of the tetragonal and cubic phases, the temperature-driven ferroelectric phase transition, and various polar topological textures in PbTiO$_3$/SrTiO$_3$ superlattices~\cite{Wu23p144102, Hu24p046802}. We have developed an online {\href{https://github.com/JiyuanY/Skyrmion-nucleus}{notebook}}~\cite{dataSkn} on GitHub to share the training database, force-field model, and training metadata. 

Unless otherwise stated, MD simulations were performed at 300~K using a time step of 2~fs using LAMMPS~\cite{Plimpton95p1}. The temperature was controlled using a Nos\'e--Hoover thermostat and the pressure using a Parrinello--Rahman barostat. A defect-free single-domain PbTiO$_3$ supercell of $16\times 16\times 50$ unit cells was initialized with its spontaneous polarization along the $+z$ direction and equilibrated for at least 100~ps before applying the electric field.
The electric fields are included in classical MD simulations using the force method~\cite{Umari02p157602}. An additional force $\mathcal{F}_i = \mathbf Z_i^{*}\cdot \mathbf{\mathcal{E}}$ was applied to each ion $i$, where $\mathbf Z_i^{*}$ is its Born effective charge tensor and $\mathbf{\mathcal{E}}$ is the applied electric field. The Born effective charges were obtained from density-functional perturbation theory using the same first-principles settings as those employed for constructing the DP training database. The electric field was applied along the $-z$ direction, opposite to the initial polarization.

\subsection*{Determination of the critical nucleus using $r$PEM}
Direct observation of three-dimensional nucleation at experimentally relevant electric fields is inaccessible to conventional MD because nucleation is rare on accessible simulation timescales. We employed a revised persistent-embryo method (\(r\)PEM) to determine the critical nucleus without prescribing its final shape. This method has previously been used to identify three-dimensional critical nuclei in the bulk, as well as two-dimensional nuclei at domain walls, during ferroelectric switching in hafnia~\cite{Yang25p021042}.

Specifically, a small reversed-polarization region containing \(N_0\) unit cells was initially embedded in the single-domain parent phase. A unit cell was classified as switched if \(P_z<-0.5P_s\). During the early stage of the simulation, a harmonic biasing potential was applied to the embryo to prevent its premature disappearance. The strength of the bias $k$ was reduced continuously as the nucleus size \(N\) increased and was completely removed once \(N\) exceeded a prescribed subcritical threshold \(N_{\rm sc}\):
\[
k(N)=
\begin{cases}
k_0 (N_{\rm sc}-N)/N, & N<N_{\rm sc},\\
0, & N\ge N_{\rm sc}.
\end{cases}
\]
Here, \(N_0<N_{\rm sc}<N^{*}\), where \(N^{*}\) is the critical nucleus size. A schematic showing the evolution of \(k\) as a function of \(N\) is presented in Fig.~S1a.

Because the bias vanishes before the system reaches the critical nucleus, configurations near the saddle point evolve on the unbiased DP energy landscape. The critical nucleus was identified from the plateau in the nucleus-size trajectory \(N(t)\), where the nucleus fluctuates between growth and shrinkage with approximately equal probability. The polarization texture and dimensions of the critical nucleus were obtained by averaging configurations sampled within this plateau region (see Fig.~S1b). Independent \(r\)PEM simulations were performed at several electric fields to determine the field dependence of the critical nucleus. We note that the accuracy of the DP model in describing supercells containing reversed domains was also comprehensively tested against DFT calculations, showing satisfactory agreement (see Fig.~S2).

\subsection*{Local polarization and topological analysis}
The local polarization of unit cell \(m\) was evaluated from the atomic displacements relative to the centrosymmetric reference structure and the corresponding Born effective charge tensors, \(\mathbf P_m = \frac{1}{V_{\rm uc}} \sum_{i\in m} w_i\mathbf Z_i^{*}\Delta\mathbf r_i\), where \(V_{\rm uc}\) is the unit-cell volume, \(\Delta\mathbf r_i\) is the displacement of ion \(i\) in unit cell \(m\) from its reference position, and \(w_i\) accounts for atoms shared by neighboring unit cells. The resulting three-dimensional local-polarization field \(\mathbf P(\mathbf r)\) was used to characterize the nucleus. The local bound-charge density was evaluated from the full polarization divergence, \(\varrho_{\rm b}(\mathbf r)=-\nabla\cdot\mathbf P(\mathbf r)\), including both in-plane and out-of-plane polarization gradients. Spatial derivatives were evaluated using finite differences on the perovskite unit-cell grid.

The topology of the polarization texture in an \(xy\) cross-section was quantified by the skyrmion number
\[
Q=
\frac{1}{4\pi}
\int
\hat{\mathbf P}\cdot
\left(
\frac{\partial\hat{\mathbf P}}{\partial x}
\times
\frac{\partial\hat{\mathbf P}}{\partial y}
\right)
\,{\rm d}x\,{\rm d}y ,
\]
where \(\hat{\mathbf P}=\mathbf P/|\mathbf P|\) is the normalized local polarization. For the central cross-section of the critical nucleus, the integration gives \(Q=-1\), identifying the nucleus as a polar skyrmion.

\subsection*{Parameters of the skyrmion-nucleus model}
Because the skyrmion-nucleus model requires the polarization and domain-wall energies at a given temperature \(T\) and external electric field \(\mathcal E\), it is necessary to account for the softening of the ferroelectric state induced by both thermal fluctuations and the applied field. We first consider thermal softening using the quartic Landau free energy \(F(P)=-\alpha_T P^2+\beta P^4\), where \(\alpha_T=\alpha_0(T_{\rm c}-T)\) and \(T_{\rm c}\) is the Curie temperature. This gives the zero-Kelvin polarization \(P_s(0)=\sqrt{\alpha_0 T_c/(2\beta)}\) and the mean-field scaling \(g(T)=\sqrt{(T_{\rm c}-T)/T_{\rm c}}\), such that \(P_s(T)=P_s(0)g(T)\). The calculated polarization value for PbTiO$_3$ is \(P_s(0)=1.0~\mathrm{C/m^{2}}\).
The local energy scale, defined as the depth of the Landau free-energy minimum, is \(K_{\rm loc}(T)=\alpha_T^2/(4\beta)=\beta [P_s(T)]^4\). The coefficient \(\beta=K_{\rm loc}(0)/[P_s(0)]^4=6.73\times10^8~\mathrm{J\,m^5\,C^{-4}}\) is temperature-independent and can be readily computed, since \(K_{\rm loc}(0)\) is essentially the energy difference between the ferroelectric and paraelectric phases of PbTiO$_3$ at zero Kelvin.

The additional softening of \(P_s(T)\), and thus of \(K_{\rm loc}(T)\), caused by an electric field antiparallel to the spontaneous polarization was then obtained from the electric enthalpy \(F(P)=-\alpha_T P^2+\beta P^4-\mathcal E P\). Expanding around the zero-field spontaneous polarization \(P_s(T)\) yields (see derivations in Supplementary Sect.~V)
\[
P(\mathcal E,T)
=
P_s(T)
\left[
1
-
\frac{\mathcal E P_s(T)}{8K_{\rm loc}(T)}
-
\frac{3}{128}
\left(
\frac{\mathcal E P_s(T)}{K_{\rm loc}(T)}
\right)^2
\right].
\]
The characteristic softening field is defined as \(\mathcal{E}^*(T)=K_{\rm loc}(T)/P_s(T)\). The same expression for the electric enthalpy is also used to determine the spinodal instability, at which the polarization becomes unstable under the opposing field. This condition is given by \(\partial F/\partial P=0\) and \(\partial^2F/\partial P^2=0\), yielding \(\mathcal{E}^{\rm max}=8K_{\rm loc}/(3\sqrt{3}\,P_s) = 8/(3\sqrt{3}\,\mathcal{E}^*) \).

The zero-field, zero-Kelvin domain-wall energies \(\sigma(0)\) for 180$^\circ$ N\'eel-type domain walls were calculated using DP. Periodic supercells containing two 180$^\circ$ N\'eel-type domain walls were constructed with wall normals along the \(x\) and \(z\) directions, respectively. The domain-wall energy was calculated as \(\sigma(0)=(E_{\rm DW}-E_{\rm SD})/(2S_{\rm DW})\), where \(E_{\rm DW}\) is the energy of the supercell containing the two domain walls, \(E_{\rm SD}\) is the energy of the corresponding single-domain supercell, and \(S_{\rm DW}\) is the area of one wall. The factor of two accounts for the two domain walls imposed by periodic boundary conditions. The calculated values are \(\sigma_x^{\rm N\acute eel}(0)=8.3~{\rm meV/\AA^2}\) and \(\sigma_z^{\rm N\acute eel}(0)=29.3~{\rm meV/\AA^2}\). The domain-wall energy at temperature \(T\) and electric field \(\mathcal{E}\) was obtained by polarization scaling using \(\sigma(P)=\sigma(0)\left[P/P_s(0)\right]^3\) (see Supplementary Sect.~VI), where \(P\) is the properly scaled value at \(T\) and \(\mathcal{E}\) as discussed above. The Curie temperature \(T_{\rm c}\) was obtained from MD simulations: constant-pressure, constant-temperature (\(NPT\)) simulations with a relaxed lattice give \(T_{\rm c}=600~\mathrm{K}\), whereas constant-volume, constant-temperature (\(NVT\)) simulations with a clamped lattice yield, as expected, a higher Curie temperature of \(T_{\rm c}=1300~\mathrm{K}\). As summarized in Table S1, The skyrmion-nucleus model only depends on four parameters, $P_s(0)$, \(\sigma_x^{\rm N\acute eel}(0)\), \(\sigma_z^{\rm N\acute eel}(0)\), and  \(T_{\rm c}\). 

\clearpage
\newpage

{\bf{Acknowledgments}} We acknowledge the support from Zhejiang Provincial Natural Science Foundation of China (LR25A040004) and National Natural Science Foundation of China (92370104). The computational resource is provided by Westlake HPC Center.

{\bf{Author Contributions}} S.L. conceived the idea and designed the project. J.Y. and D.L. performed the molecular dynamics simulations and analyzed the data. J.Y. and S.L. co-wrote the manuscript. All authors reviewed and commented on the manuscript.

{\bf{Competing Interests}} The authors declare no competing financial or non-financial interests.

{\bf{Data Availability}} The data that support the findings of this study are included in this article and are available from the corresponding author upon reasonable request.

\clearpage
\newpage 
\bibliography{SL.bib}

\clearpage
\newpage

\clearpage
\newpage

\clearpage
\newpage

\clearpage
\newpage

\end{document}